\documentclass[11pt,preprint]{aastex}

\usepackage{epsfig}
\usepackage{color}

\newcommand{\etal}{et~al. }

\newcommand{\parder}[2]{\frac{\partial{#1}}{\partial{#2}}}

\begin{document}
\title {Scalability of Hydrodynamic Simulations}
\author{\vspace{-\parskip}
Shikui Tang and Q. Daniel Wang}
\affil{\vspace{-\parskip}
  Department of Astronomy, University of Massachusetts,
  Amherst, MA 01003; tangsk@astro.umass.edu and wqd@astro.umass.edu  }

\begin{abstract}
Many hydrodynamic processes can be studied in a way that is 
scalable over a vastly relevant physical
parameter space. We systematically examine this scalability,
which has so far only briefly discussed in astrophysical literature.
We show how the scalability is limited by various constraints
imposed by physical processes and initial conditions.
Using supernova remnants in different environments
and evolutionary phases as application examples, we demonstrate the
use of the scaling as a powerful tool to explore the interdependence
among relevant parameters, based on a minimum set of simulations.
In particular, we devise a scaling scheme that can be used
to adaptively generate numerous seed remnants and plant them into 
3D hydrodynamic simulations of the supernova-dominated interstellar medium. 
\end{abstract}

\keywords{methods: miscellaneous ---
galaxies: ISM ---  ISM: structure --- supernova remnants
 }

\section{Introduction} \vspace{-\parskip}

Similar natural phenomena, which may occur on vastly different 
space and time scales, can often be treated in the same way.
One well-known example is the similarity between a supernova (SN)
in the interstellar medium (ISM) and a nuclear explosion in the
earth atmosphere despite of their vastly different
energies ($\sim 10^{51}$ ergs vs. $\sim 10^{21}$ ergs).
Their blastwave structure and evolution may be mathematically
approximated by the same self-similar Sedov-Taylor solution
\citep{Sedov59}, with an appropriate scaling according to the
energy and ambient medium density. Such self-similarity,
though often limited in its applicability (e.g., the evolution
needs to be adiabatic; the mass of the ejecta is negligible; etc.),
has been widely used in astrophysical studies.

The scalability of a hydrodynamic process, as will be demonstrated 
in the present paper, has a much broader application. Here we 
explore how the solution (or simulation) for one physical setup can 
be scaled to another, when the underlying governing equations are the same.
The self-similarity is then only a special case of the scalability. 
Therefore, the scalability analysis provides a systematic way to
examine  the physical parameter space, based on a limited number of solutions.
As a specific example, we apply our scalability analysis to the study
of the SN remnant (SNR) evolution in various environments and at different
evolutionary stages.

The scalability has the same idea as the {\it homology relations},
which are used in studying the interior structure of stars in complete
equilibrium (both hydrostatic and thermal;  e.g., \citealt{KW94}).
A unique scaling relation to study the SNR evolution was probably
first introduced by \citet{Sgro72}. \citet{Chevalier74} discussed the
same scaling relation to analyze the evolution of SNRs of different
setups with a limited number of simulations.
By recognizing that one simulation of a particular SNR can be 
used to describe a family of SNRs if they all have the same $E_{sn} n_0^2$
(where $n_0$ is the number density of ambient medium and $E_{sn}$ is the
SN energy, see \S2.2 for further discussion),
\citet{Shelton99} pointed out the usefulness
of the scaling in interpreting observations with a few simulations.
These discussions, though limited in their scope, have demonstrated
the potential of using the scalability in the study of SNRs.

In the present paper, we attempt to give a systematic examination
of the scalability of SNR solutions and simulations 
and provide specific application examples.
The initial motivation of this work is to find an
effective method to generate 1D SNR seeds that can be 
embedded into 3D hydrodynamic simulations of the SN-dominated ISM,
particularly in galactic bulges where the ISM is dominated
by diffuse hot gas. The SNR evolution in such ambient medium
in general cannot be described
by the self-similar Sedov-Taylor solution, which assumes a cool 
ambient medium (hence with no energy content).
In fact the evolution depends on both the density and temperature of
the ambient medium \citep{Tang05}. Each 3D simulation needs, for example,
more than $10^4$ SNR seeds for a bulge of an even moderate
stellar mass $\sim 10^{10}\,M_\odot$, as in our Galaxy or M31,
over a few times their dynamic time scales ($\sim 10^8$ years).
The seed embedding, worked with an adaptive mesh refinement 
scheme, can effectively extend such a 3D simulation to include
the subgrid evolution of SNRs. Here the subgrid evolution
means that the structure of SNR seeds results from the evolution on
scales much smaller than the highest spatial resolution available
in the 3D simulations. The size of an embedded SNR seed cannot be
too big (in order to use the 1D simulation properly) or too small (to be
within the limited dynamic range of a 3D simulation).  Therefore,
we should adaptively select suitable SNR seeds according to the 
local density and temperature (values and gradients) of the 
environments. For each selected SNR seed, we need the 1D radial 
density, temperature, 
and velocity profiles, with proper normalizations to guarantee
the mass, momentum, and energy conservations of the embedding 
into the 3D simulation \citep{Tang09}.
In principle, we could draw the seeds (with some
interpolations) from a library of the profiles in a grid of the three
parameters: SNR radius as well as the density and temperature of 
the ambient gas (the explosion energy and ejecta mass are assumed
to be the same for all SNe;
otherwise a larger parameter space is required for such a library).
Clearly, this approach of generating and using such a large library
is not elegant, if practical. Instead,
we find that we can use the scalability to generate the
seeds based on a very limited number of 1D SNR simulations.
We describe this simple approach as an application example.

In \S2 we show how to deduce the scaling relation
starting from the basic gas dynamics equations
and what the constraints of the scaling are.
In \S3 we apply the scaling to specific cases of the SNR
evolution. In particular, we demonstrate how we use the scaling to
generate SNR seeds for the 3D simulations with a few 1D 
simulations and how to correctly interpret the simulated relations.
Finally in \S4, we discuss the potential use of the scaling 
in a broader context.

\section{Scaling Scheme} \vspace{-\parskip}
\subsection{Basic Idea}
For a system passively evolving without source terms,
the dynamics can be described by the following equations:
\begin{eqnarray}
&\ &\parder{\rho}{t} + \nabla\cdot(\rho\mathbf{v}) = 0, \\
&\ &\parder{\rho\mathbf{v}}{t} + \nabla\cdot(\rho\mathbf{v}\mathbf{v})+
\nabla P = 0, \\
&\ &\parder{\rho e }{t} + \nabla\cdot[(\rho e + P)\mathbf{v}]= 0,
\end{eqnarray}
where $\rho$, $\,\mathbf{v}$, and $\,P$ denote density, velocity vector, and
pressure, while the total specific energy $e$ can be expressed
as $e\!\!=\!{{p}\over{(\gamma-1)\rho}}\!+\!{1\over 2}\mathbf{v}^2$ for ideal gas.
This set of equations is in a closed form; i.e.,
we can solve five independent unknown scalar variables ($\rho$, $\mathbf{v}$,
and $P$; $t$ and the spatial position are explicitly known variables),
from the five equations (because Eq.~2 can be decomposed in three
scalar equations). If the thermal state of the gas is of interest,
the equation of state is needed:
\begin{equation}
  p = R_\mu \rho T
\end{equation}
where $R_\mu$ is the equivalent ideal gas
constant\footnote{Note that $R_\mu=k/\mu m_p$,
where $k$ is the Boltzmann constant, $m_p$ is the proton mass,
and $\mu$ is the average atomic weight. The value of $\mu$ depends
on the gas ionization state and might change with temperature.
For hot gas $\mu$ only weakly depends on temperature so
$R_\mu$ can be approximated as a constant. And as long as the
scaling is within a limited temperature range, the small
variation in $\mu$ can be neglected.}
and $T$ is the gas temperature.
The scaling scheme can be found by converting each
variable\footnote{These variables include fundamental quantities such
as length ($L$), time ($t$), and mass ($M$)
and other physical quantities such as density ($\rho$),
pressure ($p$), velocity ($v$), specific energy ($e$), 
total energy ($E$), etc.}
$Q$ as $Q\lambda^{i_Q}$ (where $\lambda$ can be any positive value)
and solving them for $i_Q$ (to recover the equations before the conversion).
It can be found that the same solution holds if
\begin{eqnarray}
i_\rho - i_t = i_\rho + i_v - i_L, \\
i_\rho + i_v - i_t = i_\rho + 2 i_v - i_L = i_p - i_L, \\
i_\rho + i_e - i_t = i_\rho + i_e + i_v - i_L = i_p+ i_v - i_L, \\
i_p = i_\rho + i_T,
\end{eqnarray}
which can be simplified as
\begin{eqnarray}
i_v  & = & i_L  -  i_t, \label{eq:iva}\\
i_\rho & = & i_p - 2 i_L + 2 i_t, \\
i_e &  = & 2i_L - 2i_t, \\
i_T & = & 2i_L - 2 i_t. \label{eq:iTa}
\end{eqnarray}

Therefore, only three indices are independent in
Eqs.~(\ref{eq:iva})---(\ref{eq:iTa}). If we consider $i_L$,
$i_t$, and ${i_M}$ to be
independent, other indices can then be expressed as 
\begin{eqnarray}
i_v = i_L  -  i_t, \label{eq:ivb}\\
i_e = 2 i_L - 2 i_t, \\
i_\rho  =  i_M - 3 i_L, \label{eq:irho} \\
i_p = i_M - i_L - 2 i_t, \label{eq:iP}  \\
i_E = i_M + 2i_L - 2 i_t, \label{eq:iE} \\
i_T  =  2i_L - 2 i_t. \label{eq:iT} 
\end{eqnarray}
Eqs.~(\ref{eq:ivb})--(\ref{eq:iE}) show that
the scaling relation of the physical quantities
can be directly inferred from their dimensions
based on the basic units of mass, length, and time.
Indeed there is no constraint on the choice of $i_L$, $i_t$, or $i_M$
from the governing equations (1)--(3) for this simple case
(see also \citealt{Ryutov99} where such a property of the equations
is called Euler similarity).
But $i_T$ is restricted by $i_L$ and $i_t$ through the equation
of state. This constraint is given because
$R_\mu$ is fixed (to a number with non-vanishing
dimension hard-wired in a specific simulation),
which reduces one degree of freedom for the scaling of the pressure, density,
and temperature.
In other words, although we have four basic units for the ideal gas
hydrodynamics (i.e., mass, length, time, and temperature),
we are only able to freely change three of them
when scaling from one case to another, i.e.,
$i_M$ and two other indices from the pool of $i_L$, $i_t$, and $i_T$.

A special class of the scalability is the self-similar solution.
In this case, clearly only one solution is needed.
However, such a solution, if exists, may not be easily expressed
in an analytic form and may be applicable only asymptotically
(e.g., when the effect of the initial condition becomes negligible).
In general, one may resort to a simulation to reach the solution.
Thus it can be studied as part of the scalability problem
considered here.

\subsection{Additional Constraints}

If Eqs. (1)--(3) have source terms,
more constraints may then be placed on the scaling relation.
For example, the inclusion of the thermal conduction term,
$q=\nabla\cdot[\kappa(T)\nabla T]$, 
at the r.h.s in equation (3) requires
\begin{equation} \label{eq:cond}
  i_M - 3i_t + i_L - 3.5 i_T  =   0
\end{equation}
for non-saturated thermal conduction in which
$\kappa(T) =k_0 T^{5/2}$, where $k_0 \sim 9\times 10^{-7}
\rm erg\,cm^{-1}\,\!s^{-1}\,\!K^{-7/2}$
is the Spitzer conduction coefficient \citep{Spitzer62}.
For saturated thermal conduction ($q \propto \rho c_s^3$)
no constraint like Eq.~(\ref{eq:cond}) is required
because it does not require an extra coefficient
with a non-vanishing dimension.
The constraint from a radiative cooling term
$-n_{_H} n_e \Lambda(T)$ in the same equation
depends on the form of emissivity $\Lambda(T)$.
For optically thin primordial gas of temperature larger than
$5\times 10^6 \rm\,K$, for example, the emissivity can be
approximated as $\Lambda(T) = \Lambda_0 T^{1/2}$ where
$\Lambda_0\sim 10^{-27}\, \rm erg\, cm^6\, s^{-1}\, K^{-1/2}$
and the constraint becomes
\begin{equation} \label{eq:plasma}
i_M + 3i_t -5i_L + 0.5 i_T  =  0.
\end{equation}
In general the cooling rate $\Lambda(T)$  does not have such a
simple power law form, so the scaling relation is
\begin{equation}\label{eq:sgro}
  i_T = 0;\ \ i_L=i_t; \ \ i_M = 2i_t
\end{equation}
when combined with Eq.~(\ref{eq:iT}),
which gives the unique scaling relation adopted by
\citet{Sgro72}. This scaling relation also makes $E_{sn} n_{0}^2$
an invariant (i.e., $i_E+2i_\rho=0$)
as used in \cite{Shelton99}.

Additional physical constraints other than those from the
governing equations may need to be placed on the scalability.
For example, the scaling requires $i_E$=0 and/or $i_\rho$=0
between solutions with an identical explosion energy and/or 
ambient density.
Note that we have three degrees of freedom for all the power indices,
if the number of the constraints is less than three (i.e., at least
one index is free to change), the solution is then scalable;
otherwise the solution pertains only to a particular problem.

Implicit constraints on the scaling relation may be imposed by initial
conditions as well. Specifically, when we scale one solution
[$\rho_a(r_a,t_a)$, $T_a(r_a,t_a)$, ...] to another
[$\rho_b(r_b,t_b)$, $T_b(r_b,t_b)$, ...],
the corresponding initial condition needs to be scaled in the same way.
For example, a particular scaling relation can be determined by
specifying $i_M$, $i_L$, and $i_t$, which in turn determines $i_v$,
$i_\rho$ and other indices via Eqs.~(\ref{eq:ivb})--(\ref{eq:iT}).
This scaling relation demands that the corresponding initial
conditions should be related by
\begin{eqnarray}\label{eq:initnvab}
  \rho_b(r_b,t_{b0})=\rho_a(r_a,t_{a0})\lambda^{i_\rho},\\
  v_b(r_b,t_{b0})=v_a(r_a,t_{a0}) \lambda^{i_v},
\end{eqnarray}
and other quantities for
\begin{equation}\label{eq:initrab}
  r_b=r_a\lambda^{i_L}, \ \ t_{b0}=t_{a0}\lambda^{i_t}.
\end{equation}
It is such demands on the initial condition that often make
one problem be unique from others (limiting the scalability
of their solutions), even if all have the same governing equations and
characteristic quantities such as total energy and mass
(see \S3.5 for further discussion).
In the following we assume that the scalable solutions do
have the required initial conditions unless being explicitly
expressed otherwise.

\section{Application Examples} \vspace{-\parskip}

We use the evolution of SNRs as a simple example to
demonstrate how the above described scalability can be used.
The scalability of an SNR solution or simulation depends on
its evolutionary stage and on the properties of the ambient medium.

\subsection{Sedov-Taylor Solution} \vspace{-\parskip}
If the ejecta mass can be neglected and the ambient gas temperature
can be approximated to be zero, then the evolution of the SNR can be
described by the Sedov-Taylor solution, which depends only on the
explosion energy $E_{sn}$ and the ambient gas density $\rho_0$.
The solution can be obtained either numerically (Taylor 1950)
or analytically (\citealt{Sedov59}).
In particular the self-similar solution of the shock front,
\begin{equation}\label{eq:rsh}
  r_{sh}(t) = \xi \left(\frac{E_{sn}t^2}{\rho_0}\right)^{1/5},
\end{equation}
is widely used, where $\xi \simeq 1.15$ for ideal gas with
the specific heat ratio $\gamma$=5/3.
From the scaling point view, following Eq.~(\ref{eq:irho}) and
(\ref{eq:iE}), we have $5 i_L = i_E + 2i_t - i_\rho$
(i.e., $r\propto E^{1/5}t^{2/5}\rho^{-1/5}$), which just
the same relation shown in Eq.~(\ref{eq:rsh}). Furthermore,
for the same remnant, we have $i_E$=0 and $i_\rho$=0, hence
$i_L$=0.4$i_t$, $i_v$=$-$0.6$i_t$, $i_p$=$-$1.2$i_t$,
and other indices following Eqs.~(\ref{eq:ivb})--(\ref{eq:iT}).
It shows that for a self-similar solution all the non-zero
indices are proportional to $i_t$.
This allows the scaling from one solution at any particular time
to another.

Of course, the Sedov-Taylor solution applies only when the SN
ejecta and ambient temperature can be neglected.
Otherwise, this self-similar solution cannot be applied.
But the scaling may still be useful.

\subsection{SNRs in Hot Gas} \vspace{-\parskip}
If the ambient temperature is not negligible, a generalized formula for
the SNR shock front can be expressed as \citep{Tang05}
\begin{eqnarray}
r_{sh}(t)&=& \int_{0}^{t} c_s \left( \frac{t_c}{t}+1\right)^{3/5} dt,\\
&=& \xi\left(\frac{E_{sn} t^2}{\rho_0}\right)^{1/5} 
F\left(-\frac{3}{5},\frac{2}{5};\frac{7}{5};-\frac{t}{t_c}\right),
\label{eq:rsedt05}
\end{eqnarray}
where $c_s$ is the sound speed of the ambient medium,
$F$ is the generalized hyper-geometric function and
is equal to 1.16 when $t=t_c$ which is defined as
\begin{equation}
t_c = \left[ \left(\frac{2}{5}\xi\right)^5
  \frac{E_{sn}}{\rho_0 c_s^5} \right]^{1/3}.
\label{eq:tc}
\end{equation}
This modification accounts for the energy content of the
swept-up ambient medium.
Note that Eq.~(\ref{eq:rsedt05}) is very similar
to the Sedov-Taylor solution Eq.~(\ref{eq:rsh})
except for the modification term $F$.
When $t>t_c$, the shock front evolution significantly
deviates from the Sedov-Taylor solution.
If the temperature of the ambient medium is zero,
then $t_c \rightarrow \infty$, $F=1$,
and Eq.~(\ref{eq:rsedt05}) is the same as Eq.~(\ref{eq:rsh}).
In general, the solution in this case is no longer self-similar. 
The reason is that the evolution requires $i_E$=0, $i_\rho$=0, and $i_p$=0.
Thus all the indices are fixed to be zero.
The internal profiles change with time
and cannot be scaled from one time to another.

But the solution is still scalable between remnants evolving in different 
environment. As illustrated in the following, one simulation in a particular
environment is sufficient to infer specific SNR solutions in other environments
with different ambient density and/or temperature.

\subsection{SNR Seed Generation for 3D Simulations} \vspace{-\parskip}

The scalability has a particularly important application in planting 
SNR seeds in 3D simulations of the ISM.
Suppose that each SNR in such a 3D simulation has the same explosion
energy (i.e., $i_E$=0)\footnote{In principle,
the explosion energy can vary as well and the resultant scaling relation
can be obtained in a similar way, and we do not need to expand the
parameter space of the SNR library.
Without losing generality, however, we have assumed the canonical value
$E_{sn}=10^{51}$\,erg for Type Ia SNe.}.
For convenience, we can choose the remaining two free indices to
be the power indices of the density and temperature,
which can be directly measured in the simulation.
The scaling relation can then be simplified as
\begin{eqnarray}
i_E = 0, \label{eq:a1begin}\\
i_M = - i_T, \\
i_L = - i_T/3 -  i_\rho/3,\label{eq:a1mid} \\
i_t = -5i_T/6 - i_\rho/3,  \\
i_v = 0.5 i_T. \label{eq:a1end}
\end{eqnarray}
Therefore, we can build a library of SNR templates. Each 
consists of the radial profiles of density,
temperature, and velocity when the shock front of the SNR has a certain
radius or age. These templates can be obtained from a 1D simulation  
of an SNR evolving in a uniform ambient medium of 
density $\rho_a$ and temperature $T_a$.
Using the library and the above scaling relation, we can generate SNR seeds 
at any time and at any position of the 3D simulation. The time and position
of each SN can be realized randomly according to the Poisson statistics and the 
stellar distribution of a galactic bulge, for example.
The procedure to embed an SNR seed into the 3D simulation is as follows:
\begin{list}{}{\leftmargin=1em}
\item[]
1) At the time step just after the SN and around its position,
determine a spherical region of radius $r_{max}$,
within which $T$ and $\rho$ are sufficiently uniform
so that a 1D SNR seed is a reasonable approximation (in practice, a fraction of
the radius $r_b = \eta r_{max}$ may be used, where $\eta < 1$); \\
2) Calculate the average density $\bar{\rho}$ and (mass-weighted)
temperature $\bar{T}$ in that region, and then determine the $i_T$ and $i_\rho$:
$i_\rho = \log_\lambda (\bar{\rho}/\rho_a)$
and $i_T = \log_\lambda (\bar{T}/T_a)$,
which then determines $i_L$ (Eq.~\ref{eq:a1mid}); \\
3) Search in the library for an SNR template which has the
shock front radius of $r_a \simeq r_b\lambda^{-i_L}$ and the corresponding
SNR age $t_a$; \\
4) Wait to a future elapsing time $t_b$ of the simulation,
when $t_b \simeq t_a\lambda^{i_t}$ is just satisfied, 
and read from the library the template (which may be interpolated to account for
the difference between $t_b$ and  $t_a\lambda^{i_t}$, though not
necessary if the time step is sufficiently small, compared to the $t_b$); \\
5) Scale each profile $Q$ of the template according
to $Q\lambda^{i_Q}$; \\
6) Plant the scaled profiles into the 3D simulation by replacing 
the values within $r_b$ of the SN position (see \citealt{Tang09} for 
more details).\\
\end{list}
\vspace*{-1\baselineskip}
As the result, we have an adaptively configured SNR
seed in the 3D simulation. This seed has a dynamically self-consistent
structure expected for the SNR evolving in the local ambient medium,
which is particularly important for accurately tracing the 
SNR structures and SN ejecta (see \S3.4). We can therefore incorporate
the sub-grid evolution of the SNR into the large-scale 3D simulation, 
which optimizes the use of the computational time and enlarges the covered 
dynamical range \citep{Tang09}. It is not clear how such realism of the 
SNR seed and the adaptiveness of its planting can be realized in other
simple way (e.g., assuming a uniform thermal energy deposition or other 
arbitrary profiles).

\subsection {SN Ejecta and Scalable Initial Condition}
When an SNR is young, the mass of the SN ejecta can be
considerable. Assuming that the mass is the same for the SNRs in
the consideration, we have $i_M$=0 as well as $i_E$=0 and $i_{\rho}$=0, as 
in the previous case.
The SNR evolution is not self-similar and asymptotically approaches
the Sedov-Taylor solution only when the swept-up mass is much
greater than the ejecta mass ($M_{ej}$) and the swept-up energy
is still negligible. 
But, the solution may still be scalable from
one SNR to another.
From Eqs.~(\ref{eq:a1begin})--(\ref{eq:a1end}), we also have $i_T$=0 
and $i_L$=$i_t$=--$i_\rho$/3. This
means that one SNR evolving within an ambient medium
of density $\rho_a$ and temperature $T_a$ can be scaled to
another SNR of the density $\rho_b = \rho_a \lambda^{-3i_L}$
but of the same temperature; these two SNRs have their 
ages linked by $t_b = t_a \lambda^{i_L}$
and have the same swept-up masses and energies.

The same scheme introduced in the previous section can
also be used for embedding SNR seeds including the ejecta.
But in this case the library of SNR templates needs
to be expanded because the scaling is now only accurate for
SNRs evolving under the same ambient temperature.
We can tabulate a series of SNRs simulated for a temperature grid.
An interpolation may be used to generate any needed seed
for a particular ambient gas temperature. If the grid is sufficiently
fine, then such interpolation may not even be needed.
For example, a logarithmical grid interval of 0.02
(i.e., only 50 SNR simulations are needed to cover an order of
magnitude temperature range) would introduce an uncertainty
of  $<2\%$ in the ejecta mass, if the template with the nearest
grid temperature is used.  Such a small variation of the ejecta mass
has a negligible effect on the SNR inner structure.

As indicated in \S2.2, the scalability of an SNR solution also
requires that its initial condition (i.e., SN ejecta profiles)
to be scalable with respect to the surrounding medium.
We find that such an initial condition can be set up within the
uncertainty of SN ejecta models. We adopt the density and velocity
profiles of a post-deflagration stellar remnant of a Type Ia SN as 
proposed by \citet{DC98}:
\begin{equation}\label{eq:SNIainitprof}
  \rho(r) = \rho_s e^{1-r/r_s}, \ \ \ v(r)= v_s {r \over r_s},
\end{equation}
where $\rho_s$ and $v_s$ are the corresponding values at
the characteristic radius $r_s$. The ejecta extends to a radius $r_i$
so that
\begin{eqnarray}
  \int_0^{r_i} 4\pi r^2 \rho(r) dr = M_{ej},\label{eq:SNIaMnorm}\\
  \int_0^{r_i} 2\pi r^2 \rho(r) v(r)^2 dr = E_{SN}.\label{eq:SNIaEnorm}
\end{eqnarray}
Outside $r_i$ is the ambient gas with an assumed uniform density $\rho_a$.
To make the initial condition
scalable, we set two dimensionless parameters,
\begin{equation}
  f_i=\rho({r_i})/\rho_a,\label{eq:fi}
\end{equation}
and
\begin{equation}
  f_m =4\pi\rho_a r_i^3/M_{ej},\label{eq:fm}
\end{equation}
to be the same for all SNRs in the consideration. The four 
Eqs.~(\ref{eq:SNIaMnorm})--(\ref{eq:fm}) thus determine the four parameters:
$r_s, r_i, \rho_s$, and $v_s$. From these equations, it is also easy to 
show that 
\begin{eqnarray}
  2x^3 (e^{1/x}-1) - 2x^2 - x = (3 f_i f_m )^{-1}, \\
  v_s = \left(\frac{E_{SN}}{1.5 f_m f_i M_{ej} \beta }\right)^{1/2},
\end{eqnarray}
where $x\equiv r_s/r_i$ and 
$\beta= 24 x^3 (e^{1/x}-1) - (24x^2 + 12x + 4 + x^{-1})$.
Thus $x$ and $v_s$ depend only on the assumed constants, $f_i$ and $f_m$.
Similarly, the ratio, $\rho_s/\rho_a= f_i e^{1/x-1}$, is again the same 
for various ambient densities. Thus, we can get any desirable SNR from 
a pre-simulated template with the above 
scalable initial condition.

To make the initial free expansion a good approximation
to be described by Eq.~(\ref{eq:SNIainitprof}), we need to have
$f_m$ much less than one (e.g., $10^{-4}$ in our examinations;
no significant difference is found if $f_m$=$10^{-6}$).
The parameter $f_i$ (adopted to be 10) determines the shape of
the initial ejecta profile; a larger $f_i$ (which would result in a larger $r_s$),
for example, and would give a flatter ejecta profile
(i.e., more ejecta mass is distributed near $r_i$).
But different choices of $f_i$ (between 1 and 100)
produce negligible effects.
The same method can also be used to produce other forms of scalable
initial ejecta profiles, e.g., a power law $\rho(r) \propto r^{-n}$
(e.g., \citealt{Truelove99}), or even a uniform distribution.


\subsection {SNR Reverse Shock}

With the scalable initial condition for the SN ejecta, we can 
further study how the evolution of an SNR reverse shock depends on various
physical parameters. We demonstrate this by studying the return time of the 
reverse shock ($t_R$, i.e., when it reaches the center). Specifically, we 
examine the relation between $t_R$ and the SN ejecta mass $M_{ej}$.
In general, this relation cannot be determined in a pure analytical form,
but can be easily identified in simulations.

\citet{Ferreira2008} have examined the relation based on a series of simulations,
in which the SN ejecta is initially distributed uniformly within a
radius of 0.1\,pc and has a radial velocity increasing linearly outward.
They show $t_R \propto M_{ej} ^{3/4}$, in contrast to
$t_R \propto M_{ej}^{5/6}$ predicted by \citet{Truelove99}
from a simple dimensional analysis. \citet{Ferreira2008} suspect that
this deviation may be caused by the non-zero ambient temperature assumed for
the SNRs in their simulations. However, we find that the deviation is most
likely due to their choice of the initial ejecta distribution,
which is not scalable. Their initial condition for their $t_R-M_{ej}$
examination requires $i_E$=0, $i_\rho$=0, and
$i_R$=0 (due to the specific choice of the initial ejecta radius),
hence $i_M$=0. Therefore, each of their simulations is specific to
a particular choice of $M_{ej}$ and is not scalable to different
$M_{ej}$ value. 

\begin{figure}[hbpt]
\begin{center}
\plotone{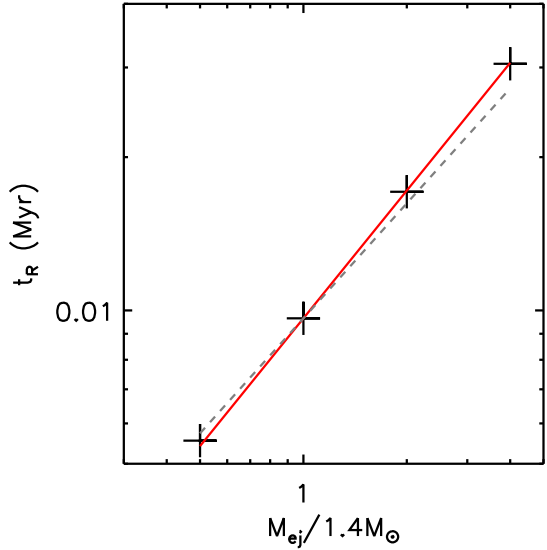}
\caption{\label{F:tr_mej}
  The relation between the reverse shock return time $t_R$
  and the ejecta mass $M_{ej}$
  for $M_{ej}=$ 0.7, 1.4, 2.8, and 5.6\,$M_{\odot}$.
  The solid line denotes the expected relation $t_R\propto M^{5/6}$,
  while the dotted line the relation $t_R\propto M^{-3/4}$
  from \citet{Ferreira2008}.
}
\end{center}
\end{figure}

Using the scalable initial condition
introduced in Eqs.~(\ref{eq:SNIainitprof})--(\ref{eq:SNIaEnorm}),
the simulations become scalable. Given $i_E$=0  and $i_\rho$=0,
it is easy to show $i_t=\frac{5}{6}i_M$ (i.e., $t_R \propto M^{5/6}$).
Based on four testing simulations with different $M_{ej}$
we identify their return times. The simulated relation of $t_R$ 
versus $M_{ej}$ is shown in Fig.~\ref{F:tr_mej}.
This result is exactly the same as the expected from the scaling relation.
The relations of $t_R$ versus $E_{sn}$ and $\rho_a$ can be
obtained similarly. Finally, we have
\begin{equation}\label{eq:tr}
  t_R \simeq 10^{4} \left(\frac{\rho_a}{m_p}\right)^{-1/3}
  \left(\frac{M_{ej}}{1.4\,\rm M_\odot}\right)^{5/6}
  \left(\frac{E_{sn}}{10^{51}\rm\, ergs}\right)^{1/2} {\rm year}.
\end{equation}
where $m_p=1.67 \times 10^{-26}
  \, \rm g\, cm^{-3}$.
The same scaling relation was also obtained for the revere shock
to reach the mantle of a core-collapse supernova exploded in
a uniform medium \citep{Reynolds84}.
\epsscale{0.46}



\section {Discussion }

We have described how a hydrodynamic solution or simulation
may be scalable and how the scalability may be used to find out 
the underlying dependence on various
physical parameters. In particular, we have demonstrated
how to apply the scaling method to adaptively generate SNR seeds 
in large-scale 3D simulations of the ISM. 
We have also discussed how an assumed initial condition
may affect the scalability, and specifically how the initial ejecta mass and
its distribution are related to the return time of the SNR
reverse shock.

Potentially, the scalability can be applied to a broad
range of topics. In the applications that we have discussed,
the ambient medium is assumed to be uniform on the relevant
scales. But the scaling method is still valid for any ambient
medium with scalable profiles such as a power law density profile
(e.g., $\rho\propto r^{-2}$ generated previously by a stellar wind).
The medium may also be clumpy. As long as the inhomogeneity does
not significantly affect the overall dynamics, which is normally true
in the early stage of SNR evolution, the scaling method may
still be applicable.

We have focused on SNRs in the hot tenuous medium for simplicity
and for the need of our practical research projects on galactic
bugles. But such SNRs are not limited to those from Type Ia SNe.
Most of SNRs from core-collapsed SNe may also evolve in hot gas
within superbubbles, because massive stars are 
born mostly in OB associations. This kind of SNRs are typically difficult
to detect (e.g., \citealt{Jiang07}),
except for situations in which radiation from pulsar wind nebulae
dominates (e.g., Crab Nebula and SNR G54.1+0.3; \citealt{Lu02}).
Luminous SNRs that are dominated by shock-heated hot gas
typically originate from run-away stars and happen to
be in a relatively dense ambient medium.
Such SNRs probably represent a minority of the entire
SNR population. In the late evolution of such an SNR,
the cooling becomes important. Even in this case, 
 the scaling method may still be useful (\S~2.2;
 \citealt{Sgro72,Chevalier74, Shelton99}).

One may also find useful applications of the scheme that we have
developed to adaptively generate SNR seeds and to embed them into 
large-scale 3D simulations of the ISM.
In particular, existing simulations of the structure formation
in the universe typically use various recipes to model the subgrid
astrophysical processes. Such recipes are often hardly calibrated with
any observations and/or are implemented in over-simplistic ways,
constrained by the limited dynamic ranges available in these simulations.
We believe that this problem may be circumvented by the application of
a scheme similar to ours, which allows for a more realistic modeling
of the subgrid evolution of important processes (e.g., individual SNRs,
superbubbles around massive stellar clusters, superwind bubbles around 
galaxies, and feedback from active galactic nuclei into the
intragroup/cluster medium). Bridging such subgrid evolution to the 
global hydrodynamics of the structure formation is badly
needed to bring the simulations closer to the reality.

\acknowledgments
We thank Bill Mathews, R. A. Chevalier, and R. Shelton for useful comments
on an early draft of this paper.
The work is supported by NASA grants NNX06AI18G and TM7-8005X (via
SAO/CXC).


\begin{thebibliography}{}
\setlength{\baselineskip}{0.0ex}
\renewcommand{\baselinestretch}{1.0}

\bibitem[Chevalier(1974)]{Chevalier74}
Chevalier R. A. 1974, ApJ, 188, 501

\bibitem[Dwarkadas \& Chevalier (1998)]{DC98}
Dwarkadas V. V., \& Chevalier R. A. 1998, ApJ, 497, 807

\bibitem[Ferreira \& de Jager(2008)]{Ferreira2008}
Ferreira S. E. S., \& de Jager O. C., 2008, A\&A, 478, 17


\bibitem[Jiang et~al.(2007)]{Jiang07}
Jiang B., Chen Y., Wang Q. D., 2007, ApJ, 670, 1142

\bibitem[Kippenhahn \& Weigert(1994)]{KW94}
Kippenhahn R., Weigert A. 1994, Stellar structure and Evolution,
corrected 3rd printing,  Springer-Verlag

\bibitem[Lu et~al.(2002)]{Lu02}
Lu F. J., Wang Q. D., Aschenbach B., Durouchoux P.,
Song L. M., 2002, ApJ, 568, L49

\bibitem[Reynolds \& Chevalier(1984)]{Reynolds84}
Reynolds S. P., \& Chevalier R. A., 1984, ApJ, 278, 630

\bibitem[Ryutov et~al.(1999)]{Ryutov99}
Ryutov D., Drake R. P., Kane J., Liang E., Remington B. A.,
Wood-Vasey W. M., 1999, ApJ, 518, 821

\bibitem[Sgro(1972)]{Sgro72}
Sgro A. 1972, Ph.D. thesis, Columbia Univ.

\bibitem[Shelton \etal(1999)]{Shelton99}
Shelton R. L., Cox D. P., Maciejewski W., Smith R. K.,
Plewa T., Pawl A., \& Rozyczka M. 1999, ApJ, 524, 192

\bibitem[Sedov (1959)]{Sedov59}
Sedov L. I 1959, Similarity and Dimensional Methods in Mechanics,
translation from 4th Russian edition, Academic press New York and London


\bibitem[Spitzer (1962)]{Spitzer62}
Spitzer L., Jr. 1962, Physics of Fully Ionized Gases
(2nd ed.; New Yor: Interscience) 

\bibitem[Tang \& Wang(2005)]{Tang05}
Tang S.,\& Wang Q. D. 2005, ApJ, 628, 205

\bibitem[Tang et al. (2009)]{Tang09}
Tang S., Wang Q. D., Mac Low M.-M., Joung M. R. 2009, astroph/arXiv0902.0386

\bibitem[Truelove \& McKee(1999)]{Truelove99}
Truelove J. K., \& McKee C. F., 1999, ApJS, 120, 299

\end{thebibliography}
\end{document}